\documentclass[11pt,a4paper]{article}
\usepackage{jheppub, amsmath, upgreek, graphicx, slashed, url, bm, hyperref}
\usepackage{color}
\usepackage{epstopdf}
\newcommand\ignore[1]{}

\newcommand{\tgr}[1]{ \textcolor{green}{\em  #1}}
\newcommand\be{\begin{equation}}
\newcommand\ee{\end{equation}}
\newcommand\bea{\begin{eqnarray}}
\newcommand\eea{\end{eqnarray}}

\title{Holographic Approach to Deep Inelastic Scattering at Small-x at High Energy}

\author[a]{Richard C. Brower}
\emailAdd{brower@bu.edu}
\author[b]{Marko Djuri\'c}
\emailAdd{djuric@fc.up.pt}
\author[c,1]{Timothy Raben\note{This work was presented at the $7^{th}$ International Conference on Quarks and Nuclear Physics 2 - 6 March 2015 Valpara\'{i}so, Chile}}
\emailAdd{timothy\_raben@brown.edu}
\author[c,2]{Chung-I Tan\note{Also presented at XXIII International Workshop on Deep-Inelastic Scattering,		27 April - May 1 2015	Dallas, Texas}}
\emailAdd{chung-i\_tan@brown.edu}
\affiliation[a]{Boston University\\
								 Boston MA 02215, USA}
\affiliation[b]{Universidade do Porto\\ 4169-007 Porto, Portugal}
\affiliation[c]{ Brown University\\ Providence, RI 02912, USA}       

\abstract{We focus on a holographic approach to DIS at small-x in high energy where scattering is dominated by exchanging a Reggeized Graviton in $AdS_5$. We emphasize the importance of confinement, which  corresponds to a deformation of $AdS_5$ geometry in the IR.  This approach  provides
an excellent  fit  to the  combined  HERA data  at small $x$. We also discuss the connection of Pomeron/Odderon intercepts in the conformal limit with anomalous dimensions in strong coupling.}

\notoc
\begin{document}
\maketitle

\newpage

\section{Introduction:}

AdS/CFT correspondence ~\cite{Maldacena:1997re, Witten:1998qj} ,  a conjectured duality between a wide class of gauge theories in d-dimensions and string theories on asymptotically $AdS_{d+1}$ product spaces, can be used to study high energy scattering processes in the non-perturbative  strong coupling limit.  It has been shown, in a holographic or AdS/CFT  dual description for QCD, the Pomeron can be identified with a reggeized Graviton in $AdS_5$~\cite{Brower:2006ea, Brower:2007xg} and, similarly, an Odderon as a reggeized anti-symmetric Kalb-Ramond $B$-field~\cite{Brower:2008cy,Avsar:2009hc}. 
 
 In the most robust example,  4 dimensional ${\cal N}=4$ Super Yang Mills theory, in the limit of large 't Hooft coupling $\lambda= g_sN_c = g^2_{ym} N_c$,  is believed  to correspond to a  limit of type IIB string theory in $d=10$. This identification partially relies on the conformal invariance of the former, but is believed to withstand deformation. The geometry on the string side is a negatively curved space times a sphere, $AdS_5\times S^5$, with Poincar\`{e} metric
\be
ds^2= \frac{R^2}{z^2} \Big[ dx^\mu dx_\mu + dz^2\Big] + R^2 d\Omega_5,
\ee
with the conformal group as its isometry. A crucial ingredient is that the dual description allows one to move from the weak to  the strong coupling region in a dual description, with the bulk coordinate $z$ serving as a length scale, ($z$ small for UV and $z$ large for IR.) 

It is important to note  that conformal invariance is broken for QCD, with a non-zero beta-function, leading to logarithmic running for $g_{ym}$ at UV and confinement in IR. Nevertheless, approximate conformal invariance remains meaningful, e.g., in perturbative treatment in the UV limit, whereas confinement  in the IR limit is often crucial in addressing non-perturbative physics. It is therefore useful to deform the metric in the bulk
\be\label{eq:AA}
ds^2\rightarrow \quad e^{2A(z)} \Big[ dx^\mu dx_\mu + dz^2\Big] + R^2 d\Omega_5
\ee
  where $A(z)\simeq -\log z$  in UV ( $z=0$)  and deviates away from the conformal limit  as $z$ increases in order to account for confinement. Under  this ``asymptotic $AdS$" setting, it is  possible to provide a unified treatment of both perturbative and non-perturbative physics at high energy.

This novel dual approach has been successfully applied to the study of HERA data~\cite{Aaron:2009wt}, both for  DIS at small-$x$~\cite{Brower:2010wf,Cornalba2010,Nishio:2011xz,Watanabe:2012uc} and for  deeply virtual Compton scattering (DVCS)~\cite{Costa:2012fw}.  More recently, this treatment has also been applied to the study of diffractive production of Higgs at LHC~\cite{Brower:2012mk} as well as other near forward scattering processes~\cite{Others}.  In this work, we first describe ``Pomeron-Graviton" duality and  its application to deep inelastic scattering (DIS) at small-x. We next turn to the issue of confinement and in particular the Pomeron singularity in a soft-wall background. We also  discuss  Pomeron and Odderon intercepts in the conformal limit and their  relation to the anomalous dimensions.  

\section{Pomeron-Graviton Duality and Holographic Treatment of DIS:}

 It can be shown for a wide range of scattering processes that the amplitude in the Regge limit, $s\gg t$, is dominated by Pomeron exchange, together with the associated s-channel screening correction, e.g., via eikonalization. In this representation, the near-forward amplitude can be expressed in terms of a Fourier transform over the 2-d transverse (impact parameter) space, 
$
A(s,t)=2i s\int d^{2}b \; e^{i \vec q \cdot \vec b} \big\{ 1- e^{i \chi(s,\vec b)}\big\} 
$ 
where $\chi(s,\vec b)$ is the eikonal. To first order in $\chi$, one has 
$
 A(s,0)\simeq 2 s\int d^{2}b \;   \chi(s,b)  \sim s^{j_0}
$
 where $j_0$ is the Pomeron intercept. 
 Traditionally, Pomeron  has been modeled at weak coupling using perturbative QCD;  
in lowest order, a bare Pomeron was first identified as a two gluon exchange, corresponding to a Regge cut in the $J$-plane at $j_0 = 1$.   Going beyond the leading order by summing generalized two gluon exchange diagrams, led to the so-called BFKL Pomeron. The position of this $J$-plane cut is at $j_0 = 1+ \log (2) \lambda /\pi^2$, recovering the two-gluon cut in the $\lambda\rightarrow 0$ limit.  In a holographic approach, the weak coupling Pomeron is replaced by the ``Regge graviton'' in AdS space, as formulated by Brower, Polchinski, Strassler and Tan (BPST)~\cite{Brower:2006ea, Brower:2007xg}.  The BPST Pomeron contains  both the hard component due to near conformality in the UV and the soft Regge component  in the IR.  To first oder in $1/\sqrt \lambda$,   the intercept moves from $j=2$, appropriate for a graviton,  down  to

\be\label{eq:BPST-intercept}
j_0 = 2 - 2 /\sqrt{\lambda}
\ee

\begin{figure}[ht]
\begin{center}
\includegraphics[width=.425\textwidth]{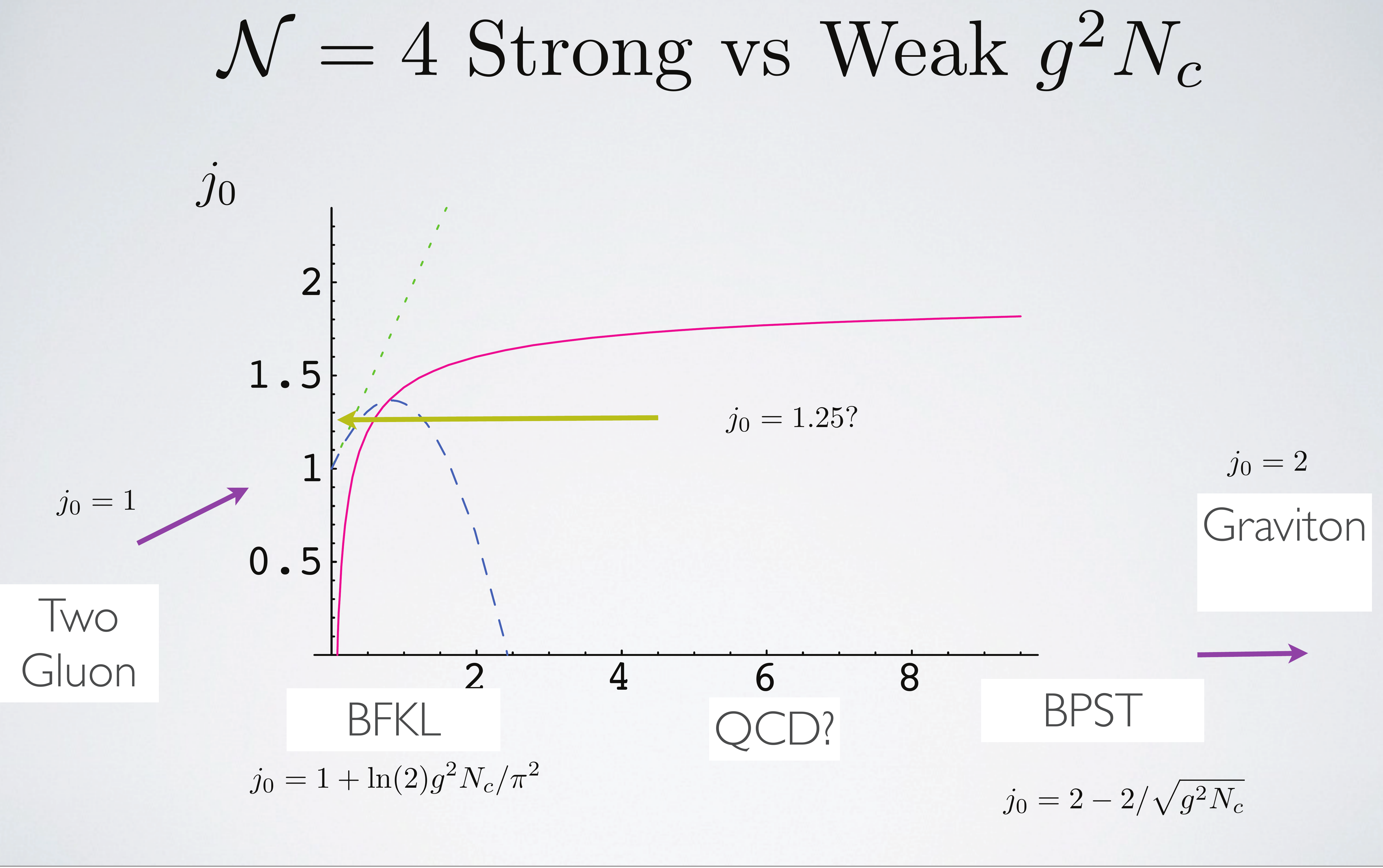}
\hskip 2.0cm 
\includegraphics[width=.425\textwidth]{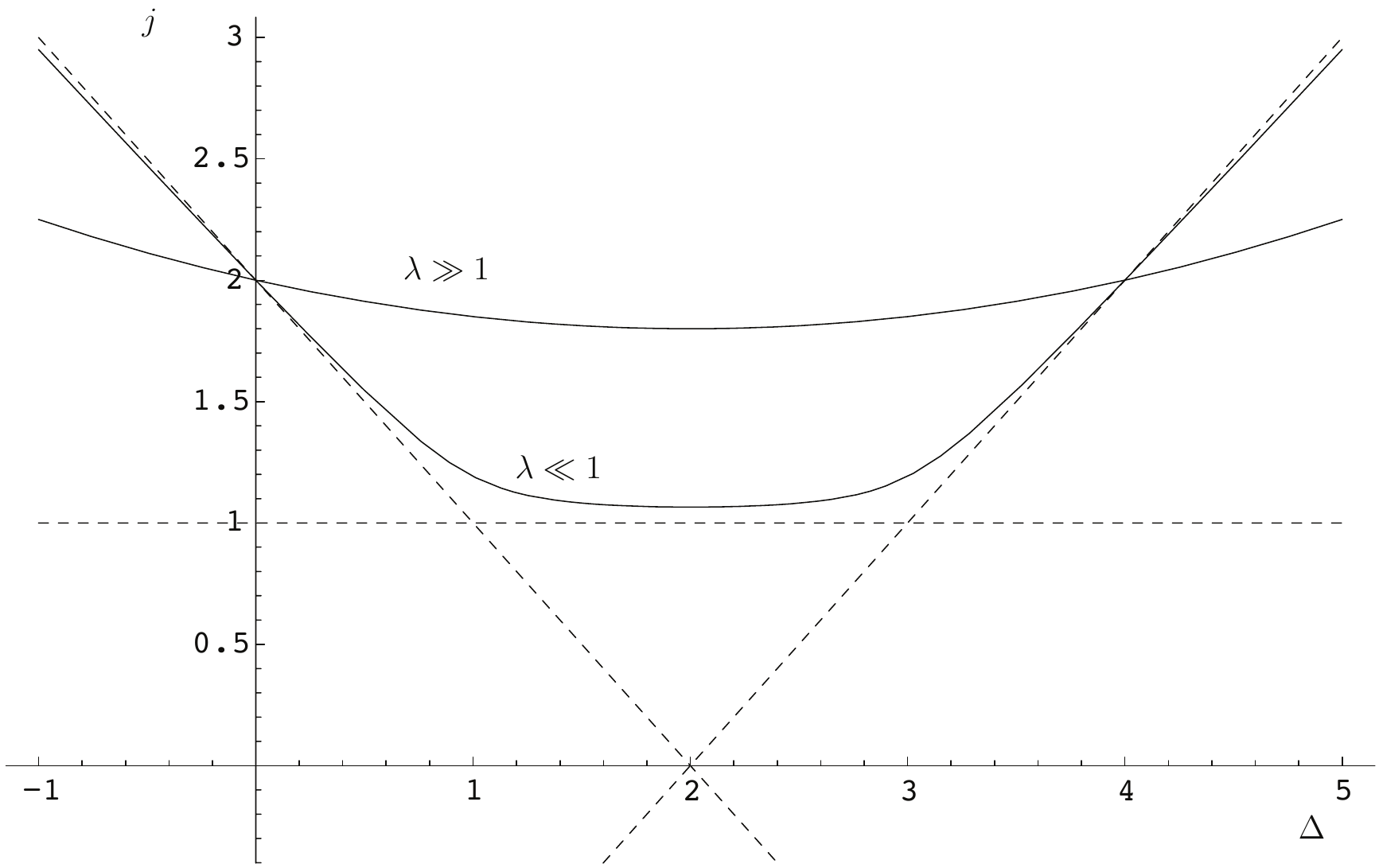}
\end{center}
\caption{ On the left,  (a), intercept  as a function of $\lambda$ for the BPST Pomeron (solid red) and for BFKL (dotted and dashed to first and second order in $\lambda$ respectively). On the right, (b), the conformal invariant $\Delta-j$ curve which controls both anomalous dimensions and the Pomeorn intercept.}
\label{fig:effective}
\end{figure}

In Fig.~\ref{fig:effective} a,  we compare the BPST Pomeron intercept   with the weak coupling BFKL intercept  for ${\cal N}=4$ YM as a function of 't Hooft coupling $\lambda$.  A typical phenomenological estimate for this parameter  for
QCD is about $j_0 \simeq 1.25$,  which suggests that the physics of diffractive scattering is in the cross over region between
strong and weak coupling.  A corresponding treatment for Odderons has also been carried out~\cite{Brower2013, Brower:2014wha}.

In a holographic approach, the transverse space $(\vec b, z)$ is 3 dimensional, where  $z \ge 0$ is the warped radial 5th dimension.     The near-forward elastic amplitude  again has the 
eikonal form~\cite{Brower:2007xg, Brower:2007qh, Cornalba:2006xm},

\begin{equation}
A(s,t)=2i s\int d^{2}b \; e^{i \vec q \cdot \vec b} \int dzdz'\; P_{13}(z)\big\{ 1- e^{i \chi(s,b,z,z')}\big\}P_{24}(z')  \; .\label{eq:A}
\end{equation}

where  $t=-q^2_{\perp}$.  For hadron-hadron scattering, $P_{ij}(z)= \sqrt{-g(z)} (z/R)^2 \phi_i(z) \phi_j(z) $  involves a product of two external normalizable wave functions for the projectile and the target respectively.
Expanding in  
$\chi(s,b,z,z')$, to first order, it is seen that the eikonal function 
is related to a BPST Pomeron kernel in a transverse $AdS_3$ representation, $\mathcal{K}({s},b,z,z') $,  with $\chi( s,b,z,z')=\frac{g_{0}^{2}}{2{s}}(\frac{R^{2}}{zz'})^2\mathcal{K}({s},b,z,z') $.

An  important unifying features for the holographic map is factorization in the AdS space.  This approach can also be applied to all DIS cross sections since they can be related to the Pomeron exchange amplitude via the optical theorem, $\sigma=s^{-1}{\rm Im} A(s,t=0)$.  For DIS, states 1 and 3 are replaced by currents, and we can simply replace $P_{13}$ by product of the  appropriate unnormalized wave-functions.  In the conformal limit, $P_{13}$ was calculated in \cite{Polchinski:2002jw} in terms of Bessel functions, so that, to obtain $F_2$, we simply  replace in (\ref{eq:A}), 
\begin{equation}
P_{13}(z)\rightarrow P_{13}(z,Q^2)=(Q^2z)[K_{0}^{2}(Qz)+K_{1}^{2}(Qz)]\,.
\end{equation}
(Similarly, for $F_1$, one has $P_{13}(z,Q^2)=(Q^2z) K_{1}^{2}(Qz)$.)  With this substitution, one has, e.g.,
\be
F_2= \frac{Q^2}{4\pi \alpha_{em}} (\sigma_T+ \sigma_L)=\frac{Q^2}{4\pi \alpha_{em} s} [{\rm Im} A(s,0)_T+ {\rm Im} A(s,0)_L]
\ee
When expanded to first order in $\chi$, Eq. (\ref{eq:A}) provides the contribution to $F_2$ from exchanging a single Pomeron, i.e., the BPST kernel,  $\mathcal{K}({s},b,z,z') $.

The momentum-space BPST  kernel in the $J$-plane, $G_j(t,z,z')$,
obeys a Schr\"odinger equation on $AdS_3$ space, with $j$ serving as eigenvalue for the Lorentz boost operators $M_{+-}$. 
 
In the conformal limit,   it takes on a simple form,
$
G_j(t,z,z')= \int_0^\infty \frac{dq^2}{2} \; \frac{J_{\tilde \Delta(j)}(zq) J_{\tilde\Delta(j)}(qz')}{q^2-t},
$
with  $\tilde\Delta(j)=\Delta(j)-2$, where    
\be
\Delta(j)=2 + \lambda^{1/4} \sqrt {2(j-j_0)}  \label{eq:Delta-j-1}
\ee
 is the conformal $\Delta-j$ curve shown in Fig.~\ref{fig:effective} b. The full Pomeron kernel can then be  obtained via an inverse Mellin transform.  In the mixed-representation, one has 
$
K(s,b,z,z')\sim -\int \frac{dj}{2\pi i} \, {\widetilde s}^j \frac{e^{-i\pi j} + 1}{\sin\pi j} \frac{e^{(2-\Delta(j))\eta}}{\sinh \eta}
\label{eq:kernel}
$
where $\cosh \eta$ is the chordal distance in $AdS_3$. By integrating over $\vec b$, one obtains  
 for the imaginary part of the Pomeron kernel at $t=0$, 
$
{\rm Im}\; \mathcal{K}(s, t=0, z,z') \sim \frac{s^{\textstyle j_0}}{\sqrt{\pi\mathcal{D}\log s}}\; e^{\textstyle -(\log z-\log z')^2/\mathcal{D}\log s} \label{eq:strongkernel}
$, 
which exhibits diffusion in the ``size" parameter $\log z$  for the exchanged closed string, analogous to the BFKL kernel  at weak coupling, with  diffusing taking place in  $\log(k_\perp)$,  the virtuality of the off-shell gluon dipole. 
The diffusion constant  takes on  $\mathcal{D} = 2/\sqrt{g^2N_c}$ at strong coupling compared to $\mathcal{D}  = 7 \zeta(3) g^2 N_c/2 \pi^2 $ in weak coupling.

\section{Fit to HERA Data:}
To confront data at HERA,  it is necessary to face the issue of confinement and saturation. Confinement can  be addressed via a hardwall cutoff, $z < z_0$,  or via a softwall model, which we shall return to shortly. The effect of saturation can next be included  via the $AdS_3$ eikonal representation (\ref{eq:A}).

To extract the key feature of holographic treatment, we shall first adopt a simplifying assumption. We note that both  integrals in $z$ and $z'$ in (\ref{eq:A}),   are  sharply peaked, the first around $z\sim 1/Q$ and the second around the inverse proton mass, $z'\equiv 1/Q'\sim 1/m_p$. To gain an understanding on the key features of dual approach, it is sufficient to   approximate both integrand by delta functions.   Under such an ``ultra-local" approximation, all structure functions take on very simple form,  e.g, 

\begin{equation}
F_{2}(x,Q^2) =\frac{g_{0}^2 }{8 \pi^{2}\lambda} \frac{Q}{Q'}\frac{e^{\textstyle (j_0 -1)\;\tau} }{\sqrt{\pi {\cal D} \tau }} \;e^{\textstyle -  (\log Q-\log Q')^2/ {\cal D} \tau}     + {\rm Confining \;\; Images}, \label{eq:f2conformalb}
\end{equation}

with  diffusion time given more precisely as $\tau =  \log ( s/QQ'\sqrt{\lambda}) =  \log (1/x) - \log (\sqrt{\lambda}  Q'/Q)$.  Here the first term is conformal. To incorporate confinement, we consider first the hardwall model where the confining effect can be expressed in terms of image charges \cite{Brower:2006ea}. It is important to note, with or without confinement, the amplitude corresponding to (\ref{eq:f2conformalb})  grows asymptotically as $(1/x)^{j_0-1} \sim s^{j_0-1}$, thus violating the Froissart unitarity bound at very high energies. The eikonal approximation in $AdS$ space, (\ref{eq:A}),  restores  unitary via multi-Pomeron shadowing~\cite{Brower:2007xg, Brower:2007qh, Cornalba:2006xm}.

\begin{figure}[ht]
\begin{center}
\includegraphics[height=0.2 \textwidth,width=.3\textwidth]{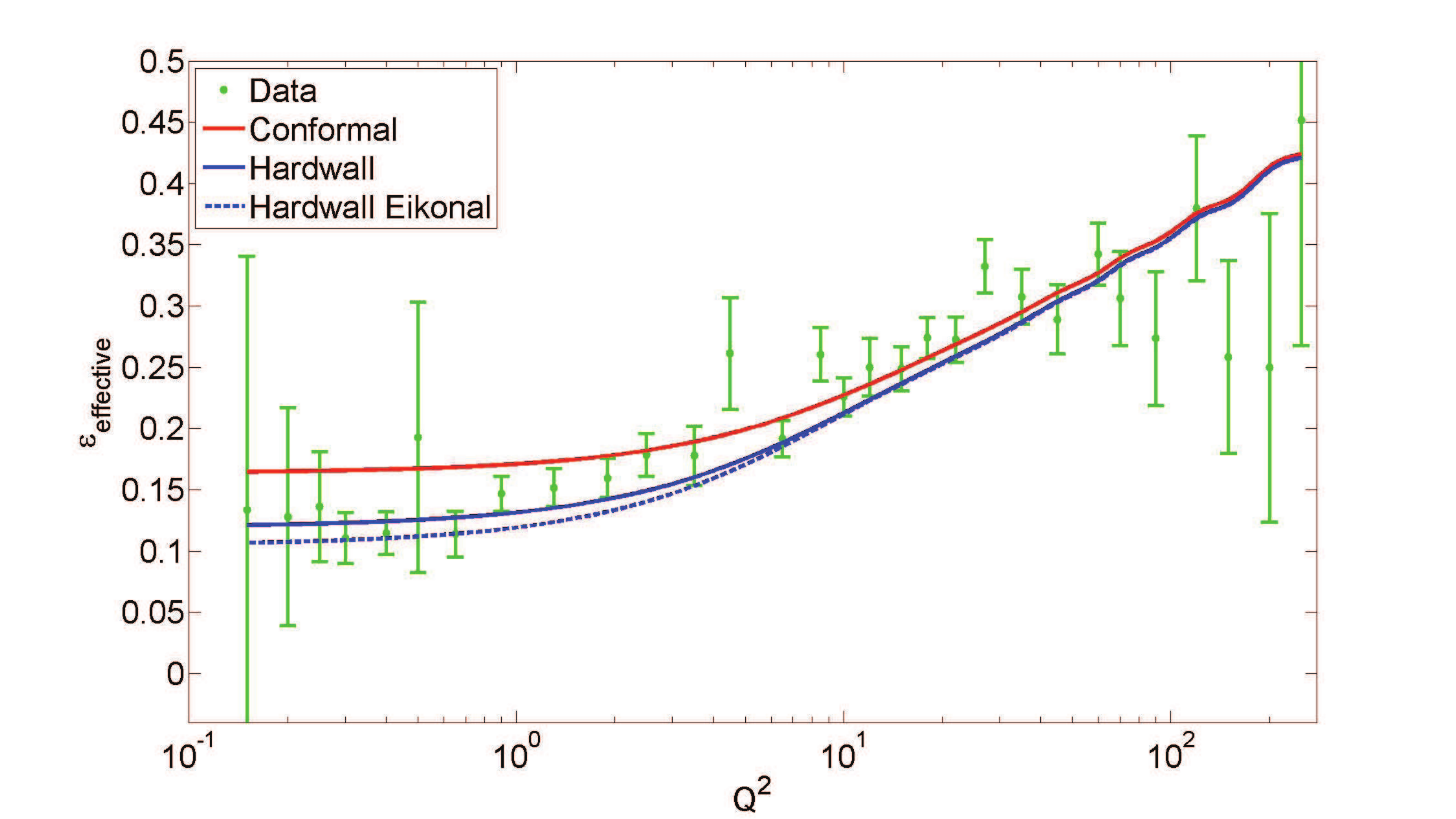}
\hfil
\includegraphics[height=0.2 \textwidth,width=.3\textwidth]{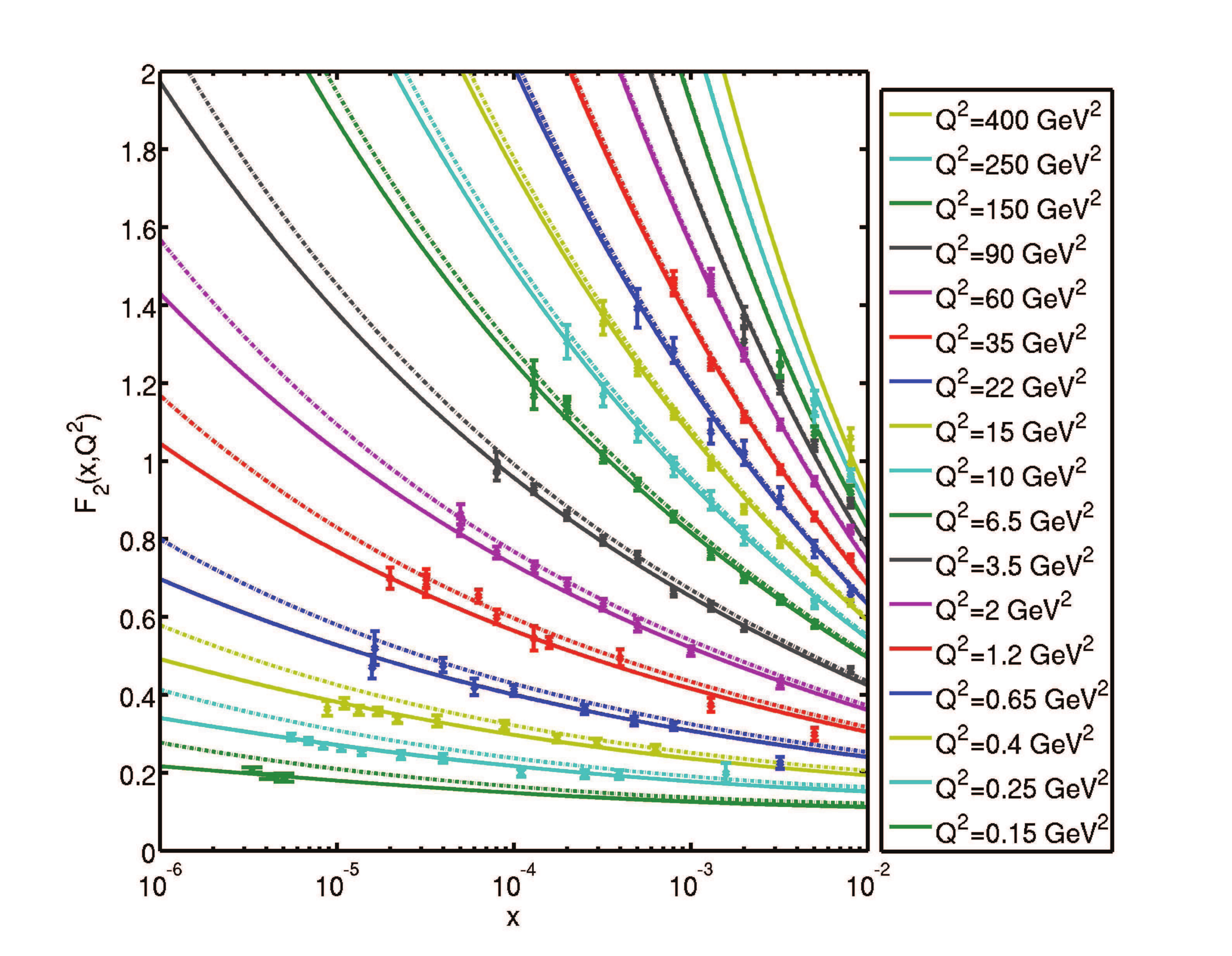}
\hfil
\includegraphics[height=0.175 \textwidth,width=.30\textwidth]{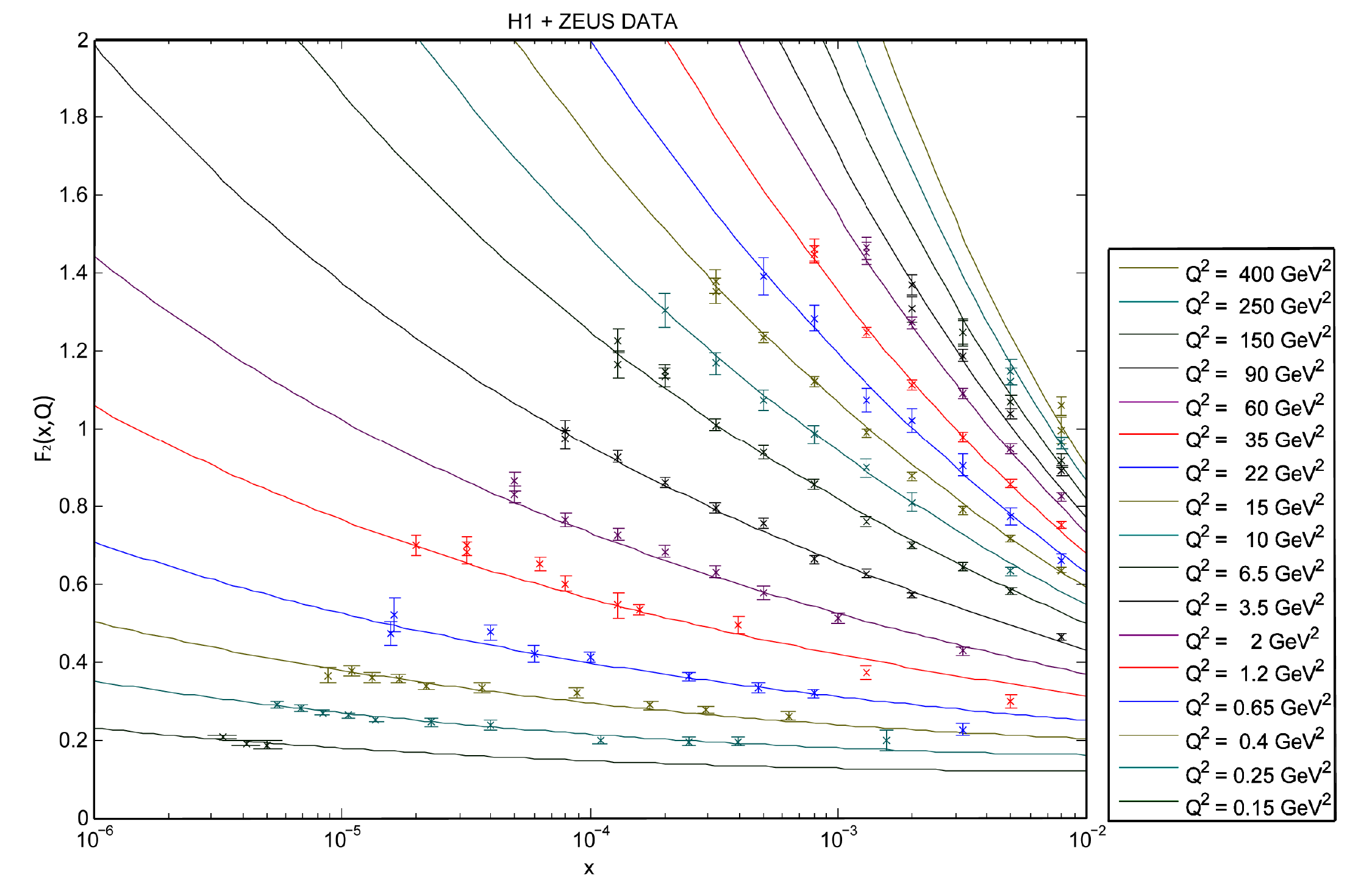}
\end{center}
\caption{In the left, (a), with the BPST Pomeron intercept at 1.22, $Q^2$ dependence for ``effective intercept" is shown for conformal, hardwall and hardwall eikonal model. In the center, (b), a more  detailed fit is presented contrasting the fits to HERA data at small x by a single hardwall Pomeron vs hardwall eikonal respectively. The softwall model, (c), was also used to fit the F$_2$ proton structure function, (c) to the right, with good success.}
\label{fig:HERALHC}
\end{figure}

We have shown various comparisons of our results~\cite{Brower:2010wf} to the small-x DIS data from the combined H1 and ZEUS experiments at HERA~\cite{Aaron:2009wt} in Fig.~\ref{fig:HERALHC}. Both the conformal, the hard-wall model, soft-wall, as well as the eikonalized hard-wall model can fit the data reasonably well. This can best be seen in Fig.~\ref{fig:HERALHC}a to the left  which exhibits the $Q^2$ dependence of an effective Pomeron intercept. This can be understood  as a consequence of diffusion. However, it is important to observe that  the hard-wall model provides a much better fit than the conformal result  for $Q^2$ less than the transition scale, $Q_c\sim 2\sim 3 $ $GeV^2$. 
The best fit to data is obtained using the hard-wall eikonal model, with a $\chi^2 = 1.04$.  This is clearly shown by Fig.~\ref{fig:HERALHC}b, where we present a comparison of the relative importance of confinement versus eikonal at the current energies.  We observe  that the   transition scale $Q_{c}^2(x)$  from conformal to confinement increases with  $1/x$, and it comes before saturation effect becomes important.  For more details, see Ref. ~\cite{Brower:2010wf}.

\begin{figure}[ht]
\begin{minipage}{0.3\linewidth}
\centerline{\includegraphics[width=0.9\linewidth]{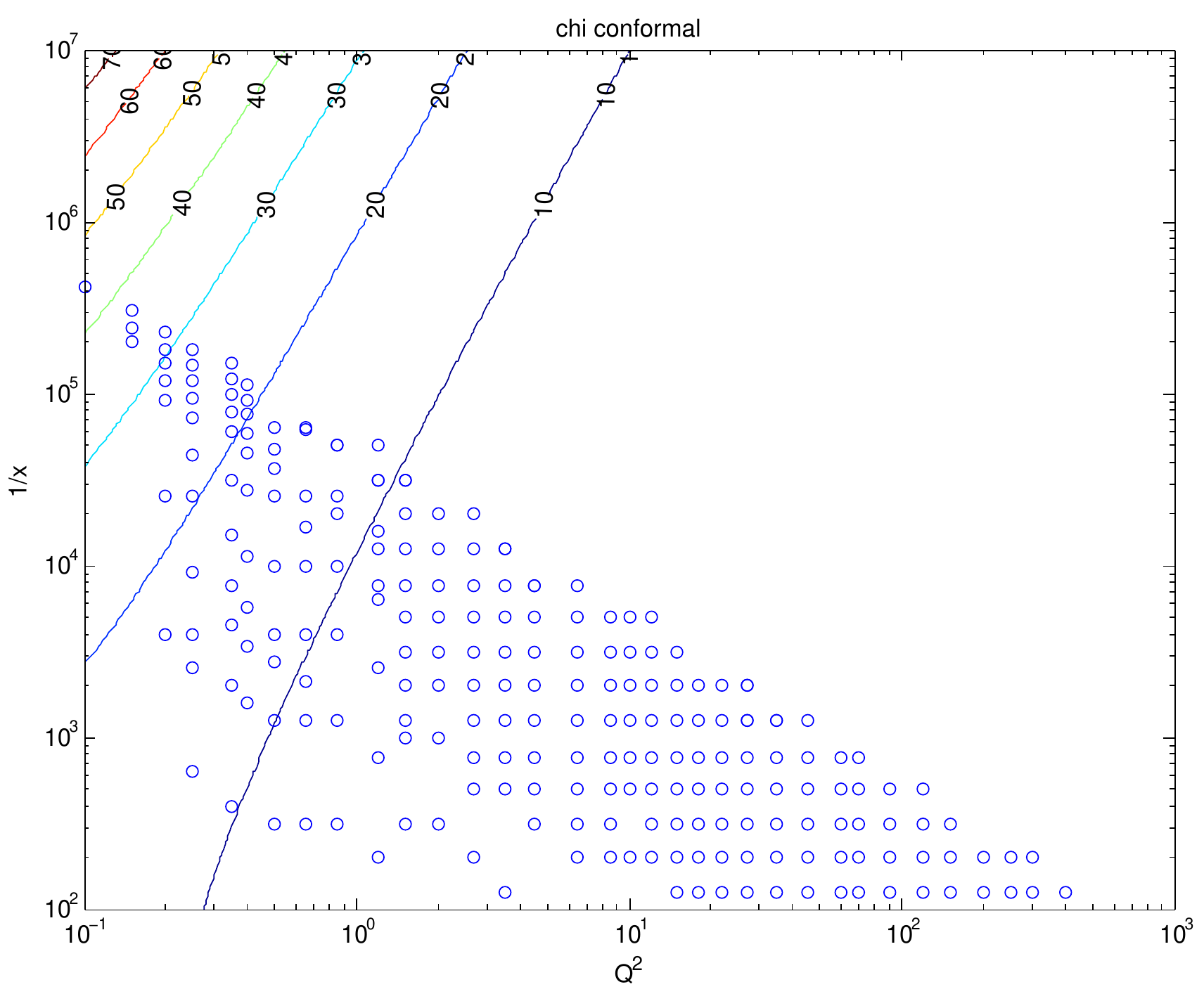}}
\end{minipage}
\begin{minipage}{0.3\linewidth}
\centerline{\includegraphics[width=0.9\linewidth]{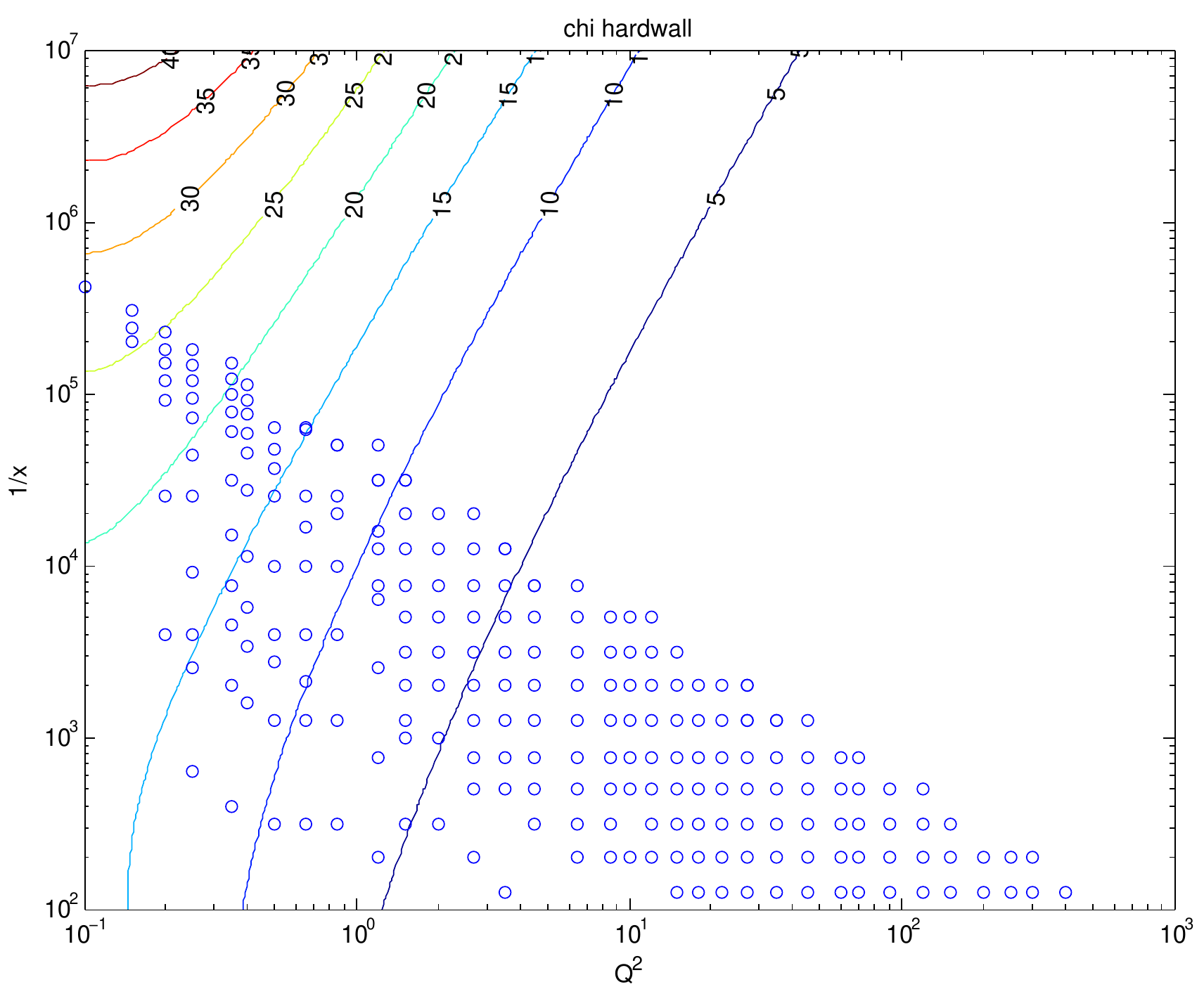}}
\end{minipage}
\begin{minipage}{0.35\linewidth}
\centerline{\includegraphics[width=0.8\linewidth]{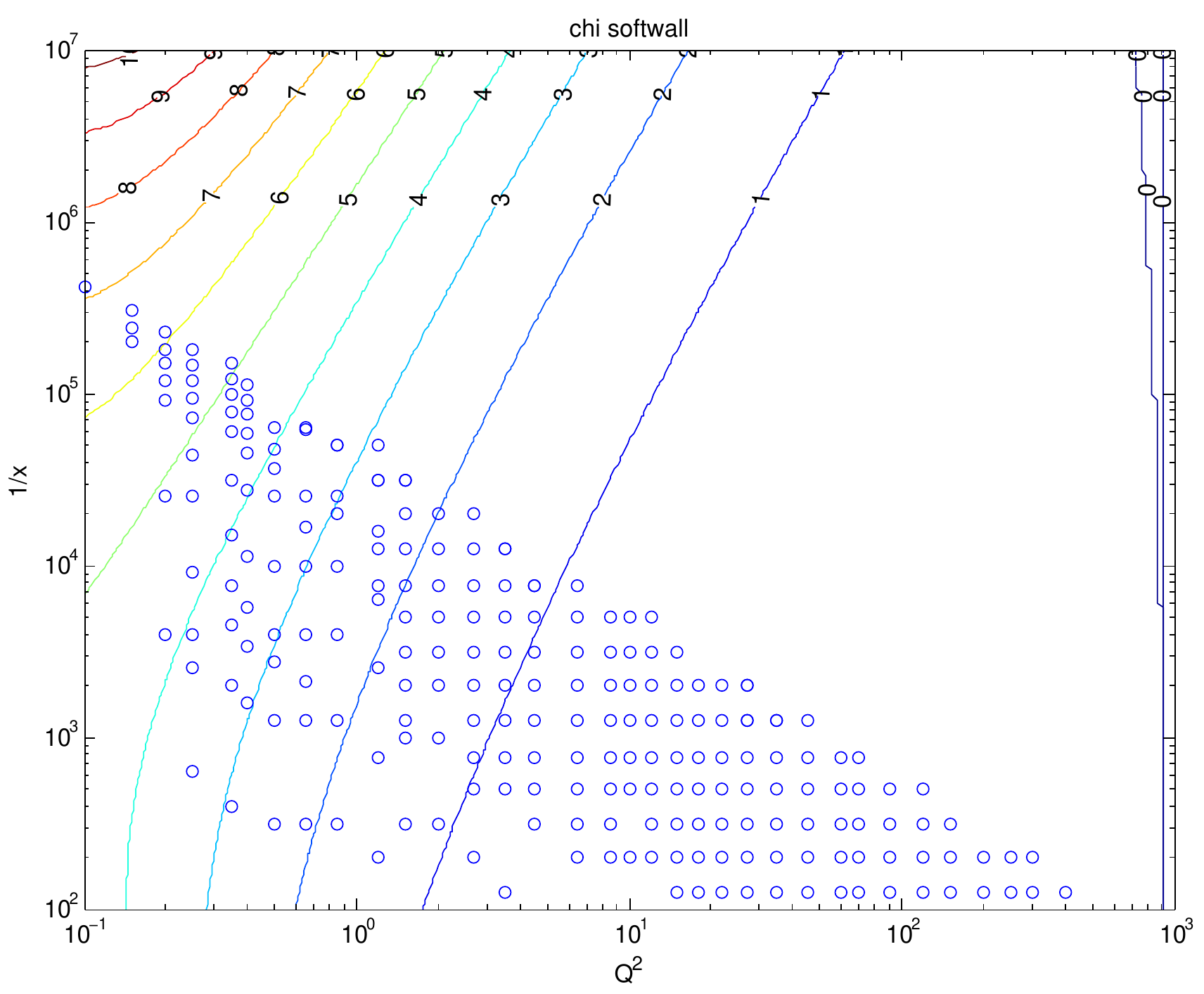}}
\end{minipage}
\caption[]{Contour plots of Im($\chi$) for the conformal (left), hardwall (center left), and softwall (center right) models.  The softwall was also used to fit the F$_2$ proton structure function (right).}
\label{fig:chi}
\end{figure}

\section{Confinement and Softwall:}

It is clear that, for $Q^2$ small, confinement is important. We find that confinement effect persists at an increasingly large value of $Q^2$ as $1/x$ increases. Equally important is the fact that
that the  transition scale $Q_{c}^2(x)$  from conformal to confinement increases with  $1/x$, and it comes before saturation effect becomes important. Therefore the physics of saturation should be discussed in a confining setting~\footnote{This has been stressed in \cite{Brower:2010wf}.  In contrast, conventional treatment, e.g., color-glass condensate, assumes that  saturation scale can be understood perturbatively.}.

The soft wall model was originally proposed in~\cite{Son}, showing what type of AdS confinement would lead to linear meson trajectories.  Several dynamical softwall toy models, where the confinement is due to a non-trivial dilaton field, have subsequently been described.~\footnote{See various references cited in \cite{Son}.  For a more general discussion of confining potentials in holographic theories and their relationship to QCD see ~\cite{Gursoy-etal} and the references therein.} There has even been some success in using the softwall model to fit QCD mesons.~\cite{Katz2006}  \footnote{These models involve an additional dynamical dilaton and tachyon field, but there is still debate about some of the signs of some parameters~\cite{Sign}}.  Significant effort has been put forth to develop standard model and QCD features in these softwall models.~\cite{Batell2008c, Csaki2007, Erlich2005}  For our purpose, as emphasized in \cite{Brower:2006ea,Brower:2007xg}, it is sufficient to consider traceless transverse  fluctuations, which do \emph{not} couple to the dilaton field, and thus a purely geometric confinement model is sufficient~
\footnote{For our present purpose, we replace $A$ in (\ref{eq:AA}) by  $c (\Lambda z)^2/3 -  \log (z/R)$, with $c=\pm 1$. The choice for $c$ remains a source of debate~\cite{Sign}. For this analysis, we shall keep $c=-1$, as originally done in \cite{Son}. It is interesting to note that the choice of the sign $c=+1$ leads to potentially intersting physics in the diffraction region of $t\leq 0$, and this and other details will be treated in a forthcoming publication.}

In the softwall model, the graviton dynamics involves a spin dependent mass-like term $\alpha^2(j)=2\sqrt{\lambda}(j-j_0)$, leading to a scalar-like equation for Pomeron propagator,  $[-\partial^2_z-(t+2c\Lambda^2)+(\Lambda^4z^2+15/4z^2)+\alpha^2(j)e^{2A(z)}/R^2]\chi_P(j,z,z',t)=\delta(z-z')$.  More precisely, for soft-wall model, each Regge contribution, label by $n$,  corresponds to a  coherent sum of exchanging t-channel modes with $t\rightarrow t_n (j) \simeq \Lambda^2(4n+d(j)$, with $n=0$ for the leading Pomeron.   the corresponding propagator can be written as combination of Whittaker's functions and their Wronskian
\be
	\chi_P(j,z,z',t)=\frac{M_{\kappa,\mu}(z_<)W_{\kappa,\mu}(z_>)}{W(M_{\kappa,\mu},W_{\kappa,\mu})}
\ee
for $\kappa=\kappa(t)$ and $\mu=\mu(j)$ .  $\Lambda$ controls the strength of the soft wall and, in the limit $\Lambda \rightarrow 0$, one recovers the conformal solution\footnote{This has a similar behavior to the weak coupling BFKL solution where Im$(\chi(p_{\perp},p_{\perp}',s))\sim\frac{s^{j_0}}{\sqrt{\pi \mathcal{D}ln(s)}}exp(-(ln (p_{\perp}')-ln(p_{\perp}))^2/\mathcal{D}ln(s))$}\hspace{5pt}$	Im(\chi_P^{conformal}(t=0))=\frac{g_0^2}{16}\sqrt{\frac{\rho^3}{\pi}}(zz')\frac{e^{(1-\rho)\tau}}{\tau^{1/2}}exp\left(\frac{-(ln(z)-ln(z'))^2}{\rho\tau}\right)$

If we look at the energy dependence of the pomeron propagator, we can see a softened behavior in the regge limit.  In the forward limit, $t=0$, the conformal amplitude scales as $-s^{\alpha_0}log^{-1/2}(s)$, but this behavior is softened to $-s^{\alpha_0}log^{-3/2}(s)$ in the hardwall and softwall models  This corresponds to the softening of of a j-plane singularity from $1/\sqrt{j-j_0}\rightarrow\sqrt{j-j_0}$.

We provide below a comparative analysis for various options. The data examined comes from the combined H1 and ZEUS experiments at HERA.~\cite{Aaron:2009wt} A fit using softwall treatment was done with the same methods used previously for the conformal and hardwall models in~\cite{Brower:2010wf}, making the results directly comparable.

\begin{table}[hb]\footnotesize
\centering
	\begin{tabular}{|c||c|c|c|c|c|}
	\hline
	Model & $\rho$ & $g_0^2$ & $z_0$ (GeV$^{-1}$) & Q' (GeV)& $\chi^2_{dof}$ \\ \hline
	conformal & $0.774^*\pm$0.0103 & $110.13^*\pm1.93$ & -- & $0.5575^*\pm0.0432$ & 11.7 $(0.75^*)$ \\ \hline
	hard wall & $0.7792\pm0.0034$ & $103.14\pm1.68$ & $4.96\pm0.14$  & $0.4333\pm0.0243$  & 1.07 $(0.69^*)$ \\ \hline
	softwall & 0.7774 & 108.3616 & 8.1798  & 0.4014  & 1.1035 \\ \hline
	softwall*& 0.6741 & 154.6671 & 8.3271  & 0.4467 & 1.1245\\ \hline
	\end{tabular}
	\caption[]{	Comparison of the best fit (including a $\chi$ sieve) values for the conformal, hard wall, and soft wall AdS models.  The final row includes the soft wall with improved intercept, calculated up to third order in $\lambda$ using Eq. (\ref{eq:P-intercept}).}
	\label{tab:fit}
\end{table}

The softwall* row describes indicates that the fit was run using a pomeron intercept (which determines $\lambda$) up to order $\mathcal{O}(\lambda^{-5/2})$~\cite{Brower2013, Brower:2014wha}.  This  has been made possible due to  integrability  and Regge techniques in $\mathcal{N}=4$ SYM.  (More on this below and also see
\cite{Basso-etal}.)  It is interesting to  examine  higher order effects further,  particularly on the confining structure in  the b-space, eikonalization, and saturation.   The will be addressed elsewhere.

For the purely conformal case, one can include saturation effects by observing when $\chi_{b=0}\sim O(1)$, at this point an eikonal approach can be used to govern multiple pomeron exchange ~\cite{Brower:2007qh}.  In the original hard wall approximation, it was shown that the $\chi$ contours are shifted to larger $Q^2$ indicating that the leading effects of confinement set in before saturation occurs.  Fig \ref{fig:chi} shows that this similar shift happens for the softwall indicating that one must address the issue of confinement before saturation sets in.  
\paragraph{Pomeron Intercept, DGLAP Connection, and Anomalous Dimensions:}
\label{sec:CFT}
 let us examine briefly the concept of  a BPST Pomerorn in more general context of conformal  field theories (CFT).  A  CFT 4-point correlation function ${\cal A}=\langle \phi(x_1) \phi(x_2)  \phi(x_3)  \phi(x_4)\rangle$ can be analyzed in an operator product expansion (OPE) by summing over allowed primary operators ${\cal O}_{k,j}$, with integral spin $j$ and dimensions  $\Delta_k(j)$,  and their descendants.  due to interaction, these conformal dimensions differ from their canonical dimension, with $\gamma_k(j) =\Delta_k(j) - j - \tau_k$, with  twist $\tau_k$.  
 
 Consider next the moments for the structure function $F_2$, $M_n(Q^2)=\int dx x^{n-2}F_2(x,Q^2)$. From OPE~\cite{Polchinski:2002jw}, or, equivalently DGLAP evolution, 
\be
M_n(Q^2)\rightarrow Q^{-\gamma(n)}
\ee
where $\gamma(n)$ is the anomalous dimension for the twist-two operators, appropriate for DIS. In particular, $\gamma_2=0$, due to energy-momentum conservation.  In our dual treatment, it is possible to identify $\gamma_n$ by our ``dimension-spin" curve, $\Delta(j)$, with $\gamma(n)= \Delta(n) - n -2$. At $j=2$, the lowest twist-2 operator  is the dimension-4 energy-momentum tensor which assures $\gamma_2=0$. 

More generally,  it was shown in ~\cite{Brower:2006ea} that $\Delta(j)$ is analytic in $j$, so that one can expand  $
 \Delta(j) $ about $j=2$ as $  \Delta(j)   = 4 + \alpha_1(\lambda) (j-2)  + O_G((j-2)^2 )$, with the coefficient $\alpha_1(\lambda)= \sqrt\lambda/4 +  O(1)$.  Equivalently, one has an expansion  
$(\Delta(j)-2)^2 = 4 + 4\alpha_1(\lambda) (j-2) + O((j-2)^2)$, consistent with Eq. (\ref{eq:Delta-j-1}), in the strong coupling limit. It was  also stressed  in \cite{Brower:2006ea} that  the $\Delta-j$ curve must be symmetric about $\Delta=2$ due to conformal invariance, and, by inverting $\Delta(j)$, one has
\be
 j(\Delta) = j(2) +\alpha_1(\lambda)^{-1}   (\Delta-2)^2 + \cdots
 \label{eq:j-Delta}
\ee
At large $\lambda$, the curve $j(\Delta)$   takes on a minimum at $\Delta=2$, as exhibited in  Fig.~\ref{fig:effective}b.   

The Pomeron intercept is simply the minimum of $j(\Delta)$ curve at $\Delta=2$, that is, $j_{0}= j(2)$. In particular, it admits an expansion in $1/\sqrt \lambda$. In a systematic expansion~\cite{Basso-etal,Brower2013, Brower:2014wha}, it has been found that
\be
\alpha_P=j_{0}=2 -\frac{2 }{\lambda^{1/2}}-\frac{1}{\lambda} +\frac{1}{ 4\lambda^{3/2}}+\frac{6\zeta(3)+2 }{\lambda^2}+\cdots,
\label{eq:P-intercept}
\ee
where terms upto $1/\lambda^{3}$ have been found. 
A similar analysis also leads to systematic expansion for the Odderon intercepts in $1/\sqrt \lambda$. As explained in \cite{Brower:2008cy, Brower2013, Brower:2014wha}, there are two odderon trajectories. One has an expansion
 $$
\alpha_O= j^{-}_0= 1 -\frac{8}{\lambda^{1/2}}-\frac{4}{\lambda} +\frac{13}{ \lambda^{3/2}}+\frac{96\zeta(3)+41}{\lambda^2}+\cdots, 
$$
where coefficients upto $1/\lambda^{3}$ have been found. Interestingly, the second trajectory, remains at $\alpha_{O,b}=1$, in dependent of $\lambda$.  
   
\section{Discussion:}

We have presented the phenomenological application of the AdS/CFT correspondence to the study of high energy diffractive scattering
for QCD.  Fits to the HERA DIS data at small x demonstrates that the strong coupling BPST Graviton/Pomerons~\cite{Brower:2006ea}  does allow for a
very good description of diffractive DIS with few phenomenological parameters, the principle one being the intercept to the bare Pomeron fit to be $j_0 \simeq   1.22$.
Encouraged by this, we plan to undertake a fuller study of several closely related diffractive process: total and elastic cross sections, DIS, virtual photon production
and double diffraction production of heavy quarks. In particular, due to explicit analytic representations of soft-wall, we can gain a better understanding on the structure of b-space in the large-b region, effects eikonalization,  and the importance of wave functions for small negative t regions. The goal is that by over constraining the basic AdS building blocks of diffractive
scattering, this framework will give 
a compelling phenomenology prediction for the double diffractive production of the Higgs in the standard model to aid in the analysis of LHC data.

\ignore{
\paragraph*{Acknowledgments:}

\tgr{\bf WE NEED THE CORRECT ACKNOWLEDGMENTS GRANTS}
}

\end{document}